\documentclass[aps,preprint,superscriptaddress,nofootinbib]{revtex4-1}   
\pdfoutput=1
\usepackage[utf8]{inputenc}
\usepackage{amssymb}
\usepackage{amsmath}
\usepackage{amsfonts}
\usepackage{graphicx}
\usepackage{color}
\usepackage{xspace}
\usepackage{hyperref}
\usepackage[normalem]{ulem}

\usepackage[section]{placeins}
\usepackage{afterpage}

\usepackage{float}
\usepackage{slashed}
\usepackage{ulem}
\usepackage{appendix}

\usepackage{cancel}

\usepackage{multirow,rotating}
\usepackage[dvipsnames]{xcolor}
\usepackage{orcidlink}   

\usepackage{subcaption}
\usepackage{graphicx}

\usepackage{changes}
\definechangesauthor[name={LXK}, color=blue]{LXK}
\begin{document}
	
      \title{Long-lived sterile neutrinos from axionlike particles at the Super Tau--Charm Facility}

      \author{Xiaokun Li\,\orcidlink{0009-0003-6486-3470}}
      \email{lixiaokun@mail.hfut.edu.cn}
      \affiliation{School of Physics, Hefei University of Technology, Hefei 230601, People’s Republic of China}

	   \author{Zeren Simon Wang\,\orcidlink{0000-0002-1483-6314}}
	   \email{wzs@hfut.edu.cn}
      \affiliation{School of Physics, Hefei University of Technology, Hefei 230601, People’s Republic of China}

	   \author{Yu Zhang\,\orcidlink{0000-0001-9415-8252}}
	   \email{dayu@hfut.edu.cn}
      \affiliation{School of Physics, Hefei University of Technology, Hefei 230601, People’s Republic of China}

      \author{Xiaorong Zhou\,\orcidlink{0000-0002-7671-7644}}
      \email{zxrong@ustc.edu.cn}
      \affiliation{University of Science and Technology of China, Hefei 230026, People’s Republic of China}

	\begin{abstract}
           We study the search prospect of long-lived heavy neutral leptons (HNLs) pair produced in decays of axionlike particles (ALPs) at the proposed Super Tau--Charm Facility (STCF), focusing on the center-of-mass energy of $\sqrt{s}=3.773$~GeV. The ALPs are assumed to originate from $D^\pm$-meson decays in association with a charged pion. We perform both truth-level and detector-level Monte Carlo simulations and obtain the expected sensitivity reach to the mixing parameter between the HNL and the electron neutrino, $|V_{eN}|^2$, with a displaced-vertex search at STCF. We find that STCF can probe values of $|V_{eN}|^2$ about one-to-two orders of magnitude beyond the existing bounds. We also perform an approximate reinterpretation of a search for HNLs at the CHARM experiment, which is subject to model-dependent assumptions on the production and kinematic distributions, and find that beam-dump experiments may provide strong complementary constraints.
 	\end{abstract}



	\maketitle
    \noindent

\section{Introduction}\label{sec:intro}

In recent years, searching for new physics in the form of long-lived particles (LLPs) has become an important direction in the field of particle physics~\cite{Alimena:2019zri,Lee:2018pag,Curtin:2018mvb,Beacham:2019nyx,Jeanty:2025wai,Alimena:2025kjv}, partly as a result of the absence of discovery of heavy, promptly decaying new fundamental particles and partly because of the rich motivations often associated with models predicting such LLPs including explaining the non-vanishing neutrino masses or dark matter.

Among the many possible theoretical scenarios where LLPs naturally arise, a class of portal-physics models are often studied, perhaps because they are representative toy models of ultraviolet completions and predict rich phenomenologies at various terrestrial experiments.
Such portals include the heavy neutral leptons (HNLs)~\cite{Shrock:1980vy,Shrock:1980ct,Shrock:1981wq} (see also Ref.~\cite{Abdullahi:2022jlv} for a review), a pseudoscalar boson such as an axion or axionlike particle (ALP)~\cite{Peccei:1977hh,Peccei:1977ur,Witten:1984dg,Conlon:2006tq,Arkani-Hamed:2006emk,Arvanitaki:2009fg,Cicoli:2012sz}, a dark scalar~\cite{OConnell:2006rsp,Wells:2008xg,Bird:2004ts,Pospelov:2007mp,Krnjaic:2015mbs,Boiarska:2019jym}, and a dark photon~\cite{Okun:1982xi,Galison:1983pa,Holdom:1985ag,Boehm:2003hm,Pospelov:2008zw,Curtin:2014cca}.
Some examples of scenarios that involve more than one of such portals are dark axion portal~\cite{Kaneta:2016wvf} that predict both an ALP and a dark photon, the hidden Abelian Higgs model~\cite{Wells:2008xg,Gopalakrishna:2008dv} with both an additional scalar and a dark photon, and an ALP-HNL portal~\cite{Berryman:2017twh,Carvajal:2017gjj,Alves:2019xpc}.

The HNLs are hypothetical spin-$\frac{1}{2}$ particles that are singlets under the Standard-Model (SM) gauge group.
They mix with the active neutrinos and are highly motivated for explaining the nonzero neutrino masses with the type-I seesaw mechanism~\cite{Minkowski:1977sc,Yanagida:1979as,Mohapatra:1979ia,Gell-Mann:1979vob,Schechter:1980gr} or various low-scale seesaw mechanisms~\cite{Mohapatra:1986aw,Mohapatra:1986bd,Akhmedov:1995ip,Malinsky:2005bi}, and serving as a dark-matter candidate~\cite{Dodelson:1993je,Shi:1998km,Dolgov:2000ew,Abazajian:2001vt,Abazajian:2001nj,Asaka:2005an,Asaka:2005pn,Asaka:2006nq} if their masses are in the keV range.
On the other hand, the QCD axion was originally proposed as a pseudo-Nambu-Goldstone boson arising from the spontaneous breaking of a global U(1)$_{\text{PQ}}$ symmetry to solve the strong CP problem~\cite{Peccei:1977hh}.
More generally, ALPs appear ubiquitously in extensions of the SM, where their masses and couplings are largely independent parameters.
They are closely connected to the strong CP problem~\cite{Peccei:1977hh}, dark matter~\cite{Preskill:1982cy,Abbott:1982af,Dine:1982ah,Arias:2012az}, and the hierarchy problem~\cite{Graham:2015cka}.
UV completions that predict the ALPs include string compactifications~\cite{Cicoli:2013ana,Ringwald:2012cu}, Froggatt-Nielsen models of flavor~\cite{Froggatt:1978nt,Alanne:2018fns}, and supersymmetry models~\cite{Bellazzini:2017neg}.

In this work, we will focus on the ALP-HNL portal and perform a phenomenological study.
In the ALP-HNL portal model, the ALPs and the HNLs can be coupled to each other, leading to phenomenologies more complicated than those arising from simple, single-portal scenarios.
Past phenomenological analyses on this model scenario have studied the ALPs produced in heavy-scalar decays~\cite{Alves:2019xpc}, electroweak processes~\cite{Marcos:2024yfm}, gluon fusion~\cite{deGiorgi:2022oks,Marcos:2024yfm,Beltran:2025oqj}, meson mixing~\cite{Abdullahi:2023gdj}, and meson decays~\cite{Wang:2024mrc,Wang:2024prt}, in a collider or beam-dump experiment setting where the ALP further decays into a pair of the HNLs; and the same new particles in the contexts of cosmology~\cite{Deppisch:2024izn,Cataldi:2024bcs} and dark matter~\cite{Gola:2021abm}.

In particular, for meson decays, one may assume tree-level quark-flavor-violating (QFV) couplings leading to the ALP production along with another meson.
Such QFV couplings are predicted in UV-completions such as the DFSZ- and KSVZ-type axion models~\cite{Ema:2016ops,Calibbi:2016hwq,Arias-Aragon:2017eww,Bjorkeroth:2018ipq,delaVega:2021ugs,DiLuzio:2023ndz,DiLuzio:2017ogq,Alonso-Alvarez:2023wig}; see also~\cite{Gavela:2019wzg,Bauer:2020jbp,Bauer:2021mvw,Chakraborty:2021wda,Bertholet:2021hjl,Bisht:2024hbs} for additional possibilities.
The quark-flavor-diagonal couplings can be suppressed, in, for example, the astrophobic axion model~\cite{DiLuzio:2017ogq} and several Froggatt-Nielsen model variants~\cite{delaVega:2021ugs,Calibbi:2016hwq,Linster:2018avp}.
In this phenomenological analysis, we choose to remain agnostic about the origin of the particular quark-flavor structure of the ALP we assume and consider the ALP QFV couplings as independent parameters.
For past phenomenological studies in this direction, see, e.g., Refs.~\cite{Gorbunov:2000ht,MartinCamalich:2020dfe,Carmona:2021seb,Bauer:2021mvw,Carmona:2022jid,Beltran:2023nli,Li:2024thq,Cheung:2024qve}.

We will study the ALPs produced in rare decays of charm mesons and then decaying promptly into a pair of long-lived HNLs, and estimate the search prospect of such long-lived HNLs at the proposed Super Tau--Charm Facility (STCF)~\cite{Achasov:2023gey,Ai:2025xop} to operate in Hefei, China.
STCF is planned to collide an electron beam and a positron beam at center-of-mass (COM) energies $\sqrt{s}$ between 2~GeV and 7~GeV with symmetric beam energy.
We will confine ourselves to the mode of $\sqrt{s}=3.773$~GeV leading to resonant production of $D^+ D^-$-pairs.
The ALPs can stem from the decays of the charged $D$-mesons in association with a charged pion and then decay promptly into a pair of long-lived HNLs.
In order to estimate the reconstruction efficiencies of the displaced vertices (DVs) arising from the decays of the long-lived HNLs, we perform both fast Monte Carlo (MC) simulations at the truth level and dedicated MC simulations at the detector level with the OSCAR framework~\cite{Huang:2022bkz,Huang:2023kog,Li:2024tuy,Ai:2024yqx} which is based on GEANT4~\cite{GEANT4:2002zbu,Allison:2016lfl} and dedicated to realistic STCF-detector simulations.
Taking into account the present bounds on the model parameters, we determine the sensitivity reach of STCF to the long-lived HNLs in terms of the mass and mixing parameters.

This paper is organized as follows.
We first introduce our model setup in Sec.~\ref{sec:model} and then provide the detail of our truth-level analysis along with a background discussion in Sec.~\ref{sec:truthlevel}.
In Sec.~\ref{sec:detectorlevel} we explain the detector-level analysis that applies within the framework of OSCAR.
The numerical results are provided in Sec.~\ref{sec:results}, and we summarize the paper in Sec.~\ref{sec:conclu}.

\section{theoretical framework}\label{sec:model}

We work within an effective framework of the axionlike particle in the following form~\cite{Bauer:2017ris,Bauer:2018uxu,Bauer:2021mvw,Carmona:2021seb,Beltran:2023nli,Alves:2019xpc,Gola:2021abm,Marcos:2024yfm}:
\begin{eqnarray}
\mathcal{L}_{a} &=& \frac{1}{2} \, \partial_\mu a  \,\partial^\mu a - \frac{1}{2}m_a^2 a^2  + \frac{\partial_\mu a}{\Lambda}  \sum_q \sum_{i,j} g^q_{i,j} \bar{q_i}\gamma^\mu q_j + \frac{\partial_\mu a}{\,\Lambda}\, g_N   \overline{N}  \gamma^\mu \gamma_5  N,
\label{eqn:Lag_ALP}
\end{eqnarray}
where $a$ is a CP-odd ALP with mass $m_a$, $N$ denotes the HNL, $\Lambda$ is the effective cutoff scale, and $q$ labels the quark fields $u_L$, $u_R$, $d_L$, and $d_R$ with generation indices $i,j=1,2,3$.

In Eq.~\eqref{eqn:Lag_ALP} the couplings $g^{q}_{i,j}$ and $g_N$ are dimensionless.
Meson decays $P\to P'/V a$, where $P$ and $P'$ denote pseudoscalar mesons and $V$ is a vector meson, can be induced if a non-vanishing coupling $g^q_{i,j}$ with $i\neq j$ is present.
In this work we confine ourselves to the flavor indices $(2,1)$ in the up-type quark sector.
With the coupling
$g^{u+}_{2,1}=g_{2,1}^{u_R}+g_{2,1}^{u_L}$ ($g^{u-}_{2,1}=g_{2,1}^{u_R}-g_{2,1}^{u_L}$), 
the charm mesons decay into an ALP plus a pseudoscalar (vector) meson. 
Specifically, a non-zero
$g^{u+}_{2,1}$ 
coupling can induce the following charm-meson decays: $D^0\to \pi^0/\eta/\eta'\,a$, $D^+\to \pi^+ \,a$, and $D_s^+\to K^+ \, a$.
On the other hand, the
$g^{u-}_{2,1}$
coupling can result in  $D^0\to \rho^0/\omega \,a$, $D^+\to \rho^+ \,a$, and $D_s^+\to K^{*+} \,a$.
We assume either $g_{2,1}^{u_L}$ or $g_{2,1}^{u_R}$ is vanishing so that both $P\to P'$ and $P\to V$ transitions occur via a single coupling, and we will label the single non-vanishing ALP-quark coupling as $g^u_{2,1}$ in the following.
We will focus on the decay $D^+\to \pi^+ a$ and its charge-conjugated channel for numerical analysis,
as our theoretical evaluations indicate that the $D^+\to \pi^+ a$ and its charge-conjugated channel serve as the dominant decay mode for ALP production.
Sensitivities can be enhanced if the other channels are included.

In the remainder of the paper, we will simply denote the coupling inducing the decay $D^+\to \pi^+ a$ with $g^u_{2,1}$.
All other ALP couplings with the quarks are assumed to be zero.

The decay rates of $D^+\to \pi^+ a$ can be computed with the following formula~\cite{Bauer:2021mvw}:
\begin{eqnarray}
 \Gamma\left(D^+ \to \pi^+ a\right) &=& \frac{|g^u_{2,1}|^2}{64\pi\Lambda^2}
 \left|F_0^{D^+ \to \pi^+}(m_a^2)\right|^2 m_{D^+}^3 \left(1 - \frac{m_{\pi^+}^2}{m_{D^+}^2}\right)^2
 \lambda^{1/2}\left(\frac{m_{\pi^+}^2}{m_{D^+}^2}, \frac{m_a^2}{m_{D^+}^2}\right)\,,\label{eqn:GammaD2Pia}
\end{eqnarray}
where $\lambda(x,y)\equiv 1+x^2+y^2-2\,x -2\,y -2xy$ and $m_{D^+/\pi^+}$ is the mass of the $D^+/\pi^+$-meson.
$F_0^{D^+\to \pi^+}$ is the $D\to \pi$ transition form factor~\cite{Wirbel:1985ji} which we extract from Ref.~\cite{Lubicz:2017syv}.
To calculate the branching ratio of $D^+ \to \pi^+ a$, Br($D^+ \to \pi^+ a$), we divide $\Gamma\left(D^+ \to \pi^+ a\right)$ by the measured total decay width of the $D^+$-meson~\cite{ParticleDataGroup:2024cfk}.

We derive the present bound on $g^u_{2,1}/\Lambda$ from the existing limits on the decay branching ratio of $D^0\to \pi^0 \nu\bar{\nu}$, since the ALP decays exclusively to a pair of long-lived HNLs in our theoretical setup which most of the time are unobserved at the main detector.
For $D^0\to \pi^0 \nu\bar{\nu}$ which is a three-body decay, we assume that the kinematics do not deviate much from those in the two-body decay $D^0\to \pi^0 a$ that would arise from the $g^u_{2,1}$ coupling.
With Eq.~\eqref{eqn:GammaD2Pia}, we derive that $g^u_{2,1}/\Lambda\lesssim 2\times 10^{-4}$~TeV$^{-1}$ (see also Ref.~\cite{Beltran:2023nli}) based on the present upper bound obtained at BESIII~\cite{BESIII:2021slf}.
In our numerical analysis, we will restrict ourselves to two representative values of $m_a$: 1.4~GeV and 1.6~GeV, and to the value of $g^u_{2,1}/\Lambda$ at the upper bound $2\times 10^{-4}$~TeV$^{-1}$.
Note that since we study HNLs pair produced in the ALP decays, the HNL mass can thus be at most 0.8~GeV in this work.

In the $m_a$ range of our interest, the ALP decays dominantly into a pair of HNLs with the decay width,
\begin{eqnarray}
    \Gamma(a\to N N)=\frac{1}{2\pi} \Big(\frac{g_N}{\Lambda}\Big)^2  m_N^2 \, m_a \sqrt{1-\frac{4 m_N^2}{m_a^2}},\label{eqn:GammaALP2NN}
\end{eqnarray}
where $m_N$ is the HNL mass.

We emphasize that in this work we assume that the HNL $N$ is of Majorana nature, and there is only one HNL kinematically relevant and it mixes solely with the electron neutrino.

For the numerical analysis, we fix $g_N/\Lambda$ at $10^{-3}$~GeV$^{-1}$ as a benchmark point, obeying the perturbativity requirement $g_N m_N/\Lambda<1$.
Thus, in the $m_a$ and $m_N$ ranges of our interest, the ALP is promptly decaying unless $m_N\approx 0$ or $m_N\approx m_a/2$
\cite{Wang:2024prt,Wang:2024mrc}.
If $g_N/\Lambda$ is orders of magnitude smaller, the ALP would become long-lived, weakening the sensitivities; see Ref.~\cite{Wang:2024mrc} where this effect was studied.
In addition, for a different value of $g_N/\Lambda$, our results remain unchanged as long as the ALP decay promptly and exclusively into a pair of the long-lived Majorana HNLs.

The HNL $N$ participates in the SM weak interactions via mixings with the active neutrinos,
\begin{eqnarray}
		\mathcal{L}_{N} &=& \frac{g}{\sqrt{2}}\ \sum_{\alpha} V_{\alpha N}\ \bar \ell_\alpha \gamma^{\mu} P_L N W^-_{L \mu} + \frac{g}{2 \cos\theta_W}\ \sum_{\alpha, i}V^{L}_{\alpha i} V_{\alpha N}^* \overline{N} \gamma^{\mu} P_L \nu_{i} Z_{\mu},
		\label{eqn:Lag_HNL}
\end{eqnarray}
where $g$ labels the SU(2) gauge coupling, $V_{\alpha N}$ is the mixing parameter between $N$ and the active neutrino $\nu_\alpha$ with $\alpha=e,\mu,\tau$, $\ell$ is an SM charged lepton, $P_L$ is the usual left-chiral projector, $W$ and $Z$ are the $W$- and $Z$-bosons, $\theta_W$ denotes the weak mixing angle, and $V^L_{\alpha i}$ is the neutrino mixing matrix in the left-handed sector between the flavor eigenstate $\nu_\alpha$ and the mass eigenstate $\nu_i$ with $i=1,2,3$.

We will work with only two parameters within the pure HNL sector: $m_N$ for the mass of $N$ and $V_{e N}$ for its mixing strength with the electron neutrino.
In the $m_N$ range of our interest, the HNL decays can be well described by two-body ones only.
Thus, for the HNL decay we need only compute the two-body decay widths at tree level, for which we follow Ref.~\cite{DeVries:2020jbs}.

We will be interested in $m_N$ ranging between 0.2~GeV and 0.8~GeV.
For this mass range, the existing bounds are led by beam-dump experiments: PIENU~\cite{PIENU:2017wbj}, KENU~\cite{Bryman:2019bjg}, CHARM~\cite{CHARM:1985nku}, NA62~\cite{NA62:2020mcv,NA62:2025csa}, T2K~\cite{T2K:2019jwa}, and BEBC~\cite{Barouki:2022bkt}.
These bounds were obtained for the HNLs in the minimal scenarios where the HNLs do not couple to any beyond-the-Standard-Model (BSM) particles.
However, for our theoretical scenario, some of these searches can apply, namely the CHARM~\cite{CHARM:1985nku} and BEBC~\cite{Barouki:2022bkt} searches.
The two searches obtained similar bounds in the mass range of our interest and we reinterpret the bounds from the CHARM search~\cite{CHARM:1985nku} in our model.
Instead of performing a full recast based on MC simulations which would be difficult, we follow a simple, quick reinterpretation method based on rescaling as proposed in Ref.~\cite{Beltran:2023nli}.
We note that this approach provides approximate results instead of precise ones.

\begin{figure}[t]
    \centering
    \includegraphics[width=0.495\textwidth]{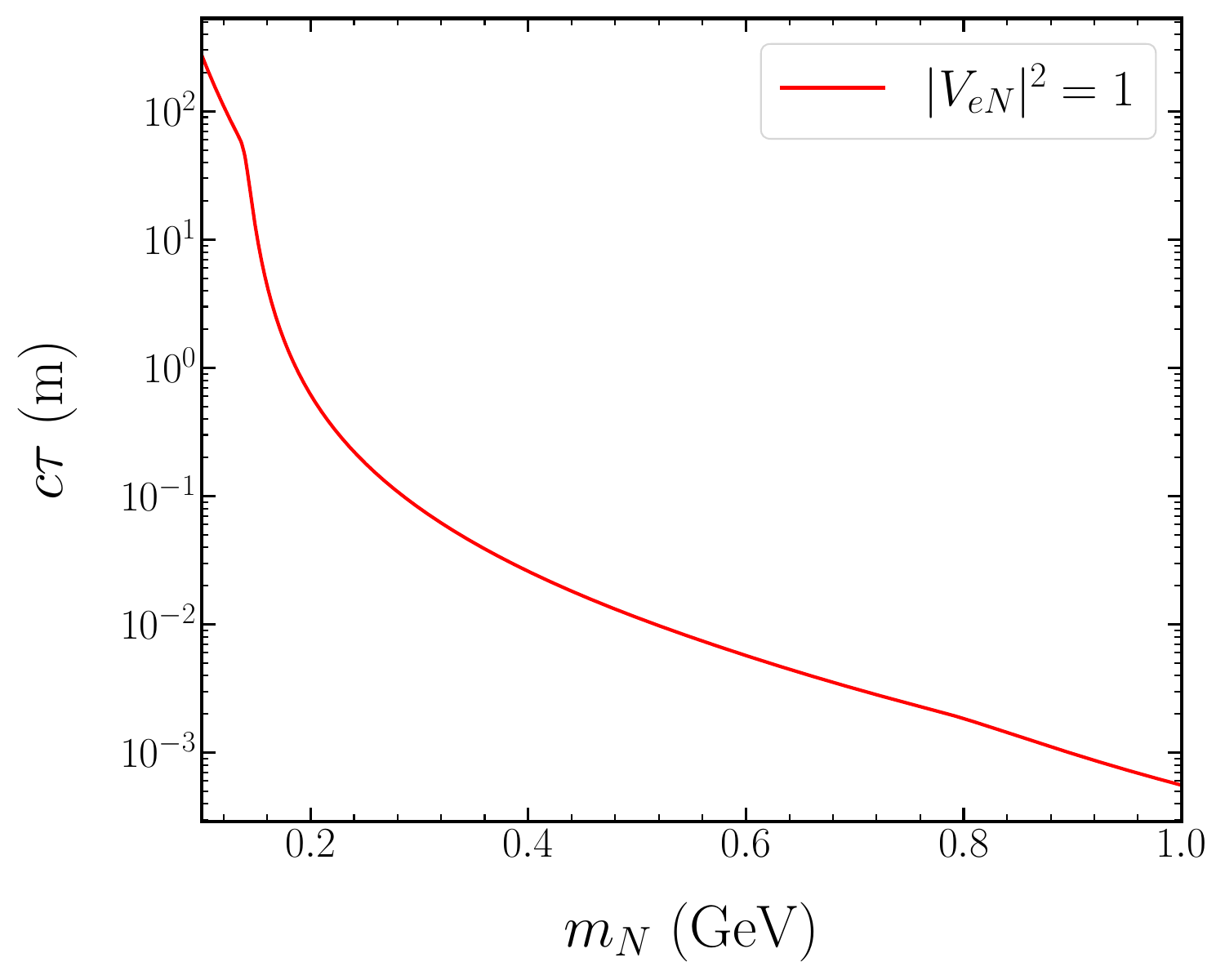}
    \includegraphics[width=0.495\textwidth]{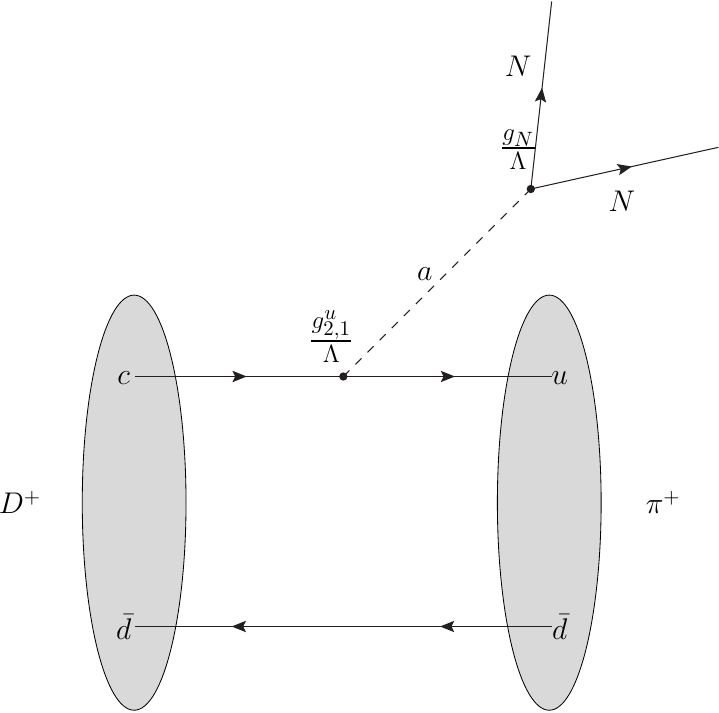}
    \caption{Left: $c\tau$ vs.~$m_N$ for $|V_{eN}|^2=1$. Right: the Feynman diagram illustrating the decays $D^+ \to \pi^+ a$ and $a\to N N$, mediated by the couplings $g^u_{2,1}/\Lambda$ and $g_N/\Lambda$, respectively.}
    \label{fig:ctau_Feynman}
\end{figure}

In Fig.~\ref{fig:ctau_Feynman} we show both a plot of $c\tau$ of the HNL as a function of $m_N$ for $|V_{eN}|^2=1$ (left) and a Feynman diagram of our signal process: $D^+\to \pi^+ a$ followed by $a\to N N$ (right).
We note that $c\tau$ is proportional to $1/|V_{eN}|^{2}$.
From the left panel, it is easy to see that the HNL can be long-lived if $m_N$ or $|V_{eN}|^2$ is small.
The right plot shows the signal process we consider; in particular, the ALP is on-shell and decays promptly.

\section{Truth-level analysis and Background discussion}\label{sec:truthlevel}

The Super Tau--Charm Facility is a next-generation $e^+e^-$ collider proposed to be constructed in China.
It is designed to achieve a peak luminosity of $0.5 \times 10^{35}\text{~cm}^{-2}\text{s}^{-1}$.
A key feature of STCF is its capability to perform energy scans over the COM-energy range of $2$--$7\text{~GeV}$, which provides extensive opportunities to search for BSM physics in the tau--charm energy region.
In particular, as a dedicated charm-meson factory operating in the resonance $\psi(3770)$, STCF is expected to produce approximately $2.8 \times 10^{9}$ $D^+D^-$ meson pairs with an integrated luminosity of $1~\text{ab}^{-1}$ collected in a single year of operation.
This yield is approximately two orders of magnitude larger than that of BESIII~\cite{BESIII:2009fln}.

STCF consists of multiple subdetectors, including the Inner Tracking system (ITK), the Main Drift Chamber (MDC), the DIRC-like Time-of-flight detector for particle identification, and the Electromagnetic Calorimeter~\cite{Achasov:2023gey}.
For charged particles in the momentum range relevant to this study (as discussed below), the identification and track reconstruction are primarily performed by the ITK and MDC.
Therefore, in the phenomenological analysis, we restrict the geometric acceptance to these two subdetectors by requiring the HNLs to decay within this region; in addition, further geometric constraints are imposed to suppress background contributions.
Specifically, the geometric coverage of the ITK and MDC can be approximated as $-140~\text{cm} < z < 140~\text{cm}$ and $6~\text{cm} < r < 85~\text{cm}$, where $z$ and $r$ denote the longitudinal and radial decay positions of the HNL, respectively.
Moreover, in the vicinity of the interaction point (IP), prompt tracks as well as soft particles and their multiple interactions with the detector material can generate a large number of detector responses, making particle identification and track reconstruction challenging in this region.
To suppress such potential backgrounds, we finally require $-140~\text{cm} < z < 140~\text{cm}$ and $10~\text{cm} < r < 85~\text{cm}$ as our fiducial volume.

\begin{figure}[t]
    \centering
    \includegraphics[width=0.495\textwidth]{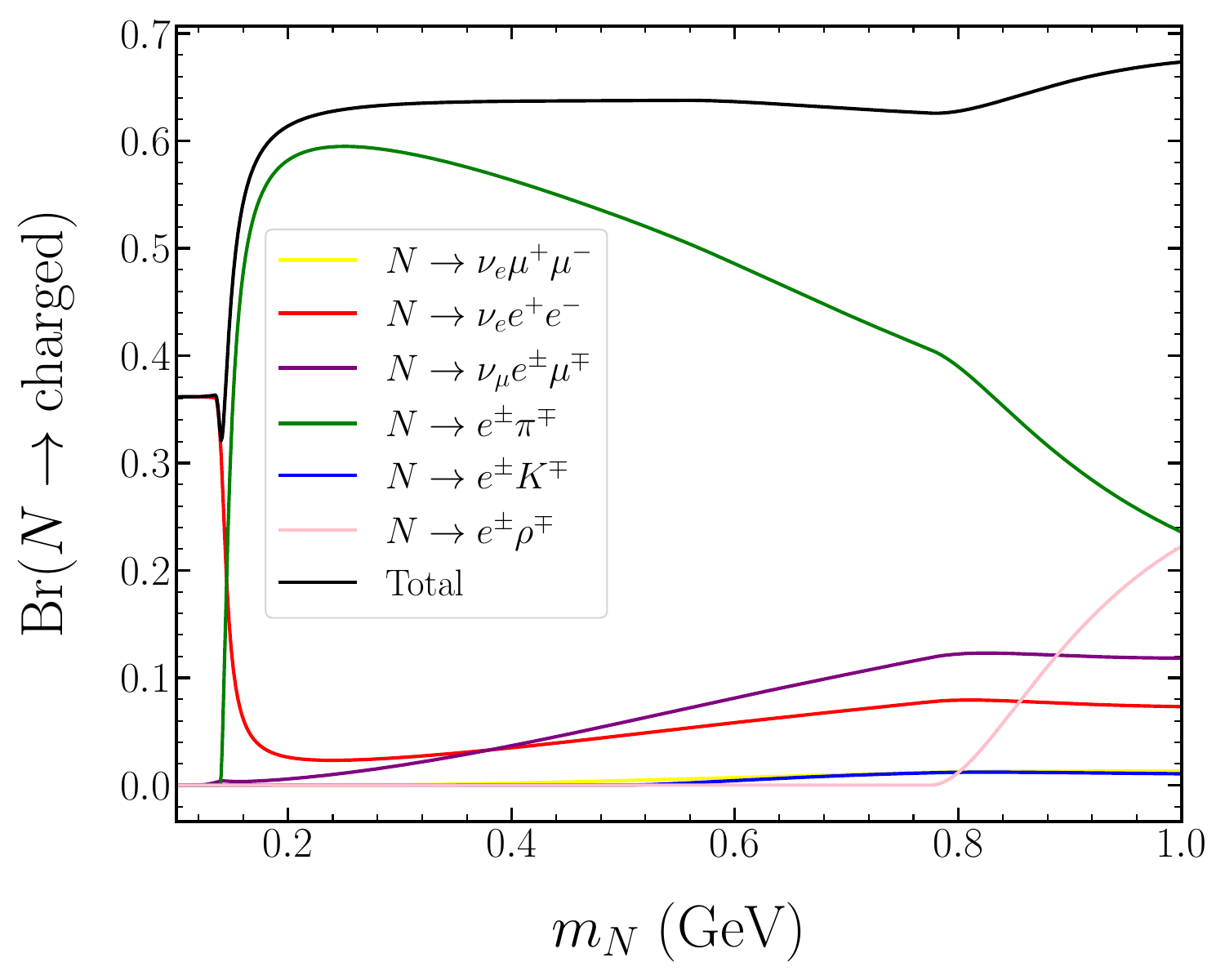} 
    \caption{The decay branching ratios of $N$ into final states including at least two charged particles, as functions of $m_N$.}
    \label{fig:visBR}
\end{figure}

Through the mixing with the electron neutrino, the HNLs can decay into a variety of possible SM final states. Generally, a higher fraction of charged particles in the HNL final states leads to increased precision of DV reconstruction and enhanced invariant-mass ($M_\text{inv}$) resolution, which improves background suppression.
Therefore, we restrict ourselves to the channels that contain at least two charged particles and do not involve significant missing energy: $\text{(1)}\ \text{leptonic}\ ll^{\prime}\nu:\ e^{\pm}\mu^{\mp}\nu_{\mu},\ e^{+}e^{-}\nu_{e},\ \mu^{+}\mu^{-}\nu_{e}$, and $\text{(2)}\ \text{semi-leptonic}\ e^{\pm}M^{\mp}:\ e^{\pm}\pi^{\mp},\ e^{\pm}\rho^{\mp},\ e^{\pm}K^{\mp}$.
We plot their branching ratios as functions of $m_N$ in Fig.~\ref{fig:visBR}.

We select the $D^-$-meson produced in the decay of $\psi(3770)$ as the tagging side and fully reconstruct it via SM decay channels (single-tag) to suppress the $q\bar{q}$ continuum background, with its kinematic properties providing constraints for the signal process on the opposite side.
On the signal side, we search for the proposed process associated with the $D^+$-meson.
Evidently, a single ALP produced on the signal side can decay into two HNLs, which may in principle both decay visibly within the detector acceptance.
In practice, the presence of a single DV is generally sufficient for effective background suppression, and the additional background rejection provided by the single tag further strengthens this.
Therefore, on the signal side, reconstructing just one HNL is considered sufficient to identify the event as a signal process.

After applying the selections given above, we expect the dominant remaining background to arise from photon conversion, which is difficult to suppress using the same strategy as that employed for backgrounds originating near the IP.
Despite imposing a stringent geometric requirement $r > 10\text{~cm}$, together with the DV constraint and the single-tag strategy, the approach of reconstructing only one HNL on the signal side remains susceptible to contamination from photon-conversion backgrounds.
Therefore, we exclude all leptonic channels including $e^{+}e^{-}\nu_{e},\ \mu^{+}\mu^{-}\nu_{e}$ and $e^{\pm}\mu^{\mp}\nu_{\mu}$, as they can be directly affected by misreconstruction of lepton pairs originating from photon-conversion processes ($\gamma \to e^+e^- / \mu^+\mu^-$).

For the STCF COM energy considered here ($\sqrt{s}=3.773\text{~GeV}$), taking into account initial-state radiation, the produced $D^+$- and $D^-$-mesons have energies approximately in the range $1.870$--$1.885\text{~GeV}$.
The momentum of the $\pi^+$-meson from $D^+$-meson decays ($D^+ \to \pi^+ a$) is required to be at least $0.1\text{~GeV}$ in order to penetrate a sufficient number of layers in the MDC and produce reconstructable tracks.
Therefore, we require the ALP mass to be below $1.6\text{~GeV}$, which further implies that the HNL mass must be less than half of the ALP mass, specifically below $0.8\text{~GeV}$.
Furthermore, an ALP with a mass $m_a$ lower than $1.0$~GeV would make the HNLs pair-produced from ALP decays kinematically incapable of decaying into $e^{\pm}K^{\mp}$, a scenario that we intend to avoid. Therefore, we constrain the $m_a$ range to $1.0$--$1.6$~GeV. Specifically, the upper bound of $1.6$~GeV and a slightly lower value of $1.4$~GeV are selected as benchmark points.
In addition, to avoid excessively soft charged particles in the HNL decay final states that would degrade the reconstruction efficiency, we impose a lower bound on the HNL mass of $0.2\text{~GeV}$.
Consequently, the range of $m_N$ is set to $0.2$--$0.8$\text{~GeV}.

In Fig.~\ref{fig:visBR} we observe that near $0.8\text{~GeV}$, the opening of the $e^{\pm}\rho^{\mp}$ channel leads to a turning point in the total branching ratios, while the branching ratio of the $e^{\pm}\pi^{\mp}$ channel begins to decrease more rapidly.
Within the HNL mass range of interest $0.2$--$0.8\text{~GeV}$, the $e^{\pm}\pi^{\mp}$ channel clearly dominates among the semi-leptonic final states; therefore, for semi-leptonic modes, we select only the $e^{\pm}\pi^{\mp}$ channel as the signal final state of the HNL.
Accordingly, the final-state particle combination considered in this work is $\pi^+ e^{\pm}\pi^{\mp}$. Its physical background can only arise from $D^{\pm} \to \pi^+\pi^- e^{\pm} \nu_e$, which can be readily rejected by the displaced-vertex requirement, as discussed above, rendering this a very clean final state. We anticipate that in a realistic detector environment, additional backgrounds such as combinatorial backgrounds or fake-track backgrounds may still be present, which will be discussed in the next section.

In addition, the MDC of the STCF detector adopts a superlayer wire configuration similar to those of Belle II and BESIII.
Such a structure causes the detector efficiency for reconstructing charged-particle tracks and particle identification (PID) to gradually decrease as the particle production point moves away from the IP, with this effect being particularly pronounced for increasing radial production positions $r$.
With a phenomenological analysis, to account for this decrease in detector efficiency, we model it using a linear function as
\begin{equation}
L(r) =
\begin{cases}
1 - \dfrac{r}{85\,\text{cm}}, & 10\text{~cm} < r < 85\text{~cm}, \\
0, & \text{otherwise}.
\end{cases}
\label{eqn:linear_efficiency}
\end{equation}
This approach has been employed in Refs.~\cite{Dib:2019tuj,Dey:2020juy,Cheung:2021mol,Bertholet:2021hjl,Cheung:2024oxh,Wang:2024prt}.
In the next section, we will present an improved, dedicated detector-level analysis for estimating the relevant effects.

In the phenomenological analysis, we implement a fast MC simulation program with phase-space decays to generate truth-level kinematic information and evaluate the signal acceptance.
Specifically, the kinematic properties of the HNL are generated through a cascade decay chain and boosts into the laboratory frame.
The decay probability is then determined according to its proper decay length ($c\tau$) to assess whether it decays within the fiducial volume.
Based on this procedure, we scan the HNL mass in the range $0.2$--$0.8\text{~GeV}$ and its proper decay length $c\tau$ in the range $10^{-6}$--$10^{10}\text{~m}$.
The final expression for the expected number of signal events $N_s$, incorporating both geometric acceptance and the linear efficiency function $L(r)$, is given by
\begin{equation}
\begin{split}
N_s
= {} & 2 \cdot N_{D^+D^-}
\cdot \sum_i \mathrm{Br}^{\,i}_{\rm tag} \cdot \epsilon^{\,i}_{\rm tag} \\
& \cdot \mathrm{Br}(D^+ \to \pi^+ a)
\cdot \mathrm{Br}(a \to NN)
\cdot \mathrm{Br}(N \to e^{\pm}\pi^{\mp})\cdot \epsilon_{\text{1DV}},
\end{split}
\label{eqn:NS}
\end{equation}
where 
\begin{eqnarray}
    \epsilon_{\text{1DV}}&=&\frac{1}{N_{\text{MC}}}\sum_{i=1}^{N_{\text{MC}}}\Big( P^{\text{1DV}}_{N_{i,1}}\cdot (1-P^{\text{1DV}}_{N_{i,2}})  + P^{\text{1DV}}_{N_{i,2}}\cdot (1-P^{\text{1DV}}_{N_{i,1}})  \Big) \label{eqn:epsilon}
\end{eqnarray}
\begin{eqnarray}
    P^{\text{1DV}}_{N_{i,j}}&=&\frac{1}{R_r^{N_{i,j}}}\int_{0}^{85\text{cm}} L(r)\,A(z_{N_{i,j}})\, e^{-r/R_r^{N_{i,j}}}\, dr. \label{eqn:prob}
\end{eqnarray}
Here, $N_{D^+D^-}$ is the total number of $D^+D^-$ pairs produced under an integrated luminosity of $1~\mathrm{ab}^{-1}$, and $\sum_i \mathrm{Br}^{\,i}_{\rm tag} \cdot \epsilon^{\,i}_{\rm tag}$ represents the sum over all SM decay channels suitable for fully reconstructing the tagging-side $D^-$-meson, each weighted by its corresponding single-tag efficiency.
The factor 2 takes into account the fact that there are two $D$-mesons in each signal event.
Using the single-tag $D^+D^-$ production events reported in Ref.~\cite{Achasov:2023gey}, we estimate a total tagging efficiency of approximately $19.64\%$.
$\mathrm{Br}(a \to NN)=100\%$.
$N_{\text{MC}}$ in Eq.~\eqref{eqn:epsilon} is the total number of the MC-simulated events.
The final integral in Eq.~\eqref{eqn:prob} simultaneously incorporates the linear efficiency function $L(r)$ and the decay probability within the fiducial volume.
$R_r^{N_{i,j}} = \beta_r^{N_{i,j}} \gamma c\tau =\frac{p_T^{N_{i,j}}}{m_N} c \tau$ is the HNL transverse decay length in the laboratory frame, with $j=1,2$ denoting the two HNLs produced from the ALP decay, and $z_{N_{i,j}}=r \cot \theta_{N_{i,j}}$ is the longitudinal decay position of the HNL, with $r$ representing the transverse decay position and $\theta$ the HNL polar angle, respectively.
The function $A(z)$ is introduced to enforce the longitudinal fiducial-volume constraint, defined as
\begin{equation}
A(z)=
\begin{cases}
1, & -140\text{~cm} < z < 140\text{~cm}, \\
0, & \text{otherwise}.
\end{cases}
\end{equation}

\section{Detector-level Analysis with OSCAR}\label{sec:detectorlevel}

To more accurately simulate the dependence of the detector efficiency on the radial production position $r$ presented in Sec.~\ref{sec:truthlevel}, we employ the SNiPER-based OSCAR framework, which is specifically developed for STCF offline data analysis.
OSCAR allows to generate physical events, simulate detector responses, reconstruct events, and analyze the signal process.

During the event generation and simulation phase, we employ the KKMC generator~\cite{Jadach:1999vf,Jadach:2000ir} to simulate $e^+e^-$ collisions and produce the $\psi(3770)$ resonance.
We use the StcfEvtGen module to handle the production and decay processes of charm mesons in both the signal and tag channels.

In the phenomenological estimation of the total tagging efficiency discussed in the previous section, all the known decay channels of the $D^-$-mesons listed in Ref.~\cite{ParticleDataGroup:2024cfk} are taken into account.
At the reconstructed level, we take the channel $D^- \to K^+ \pi^- \pi^-$ as a representative benchmark channel for evaluating the reconstruction efficiencies of the signal channel, motivated by its relatively large branching ratio and clean charged final states.
Assuming the signal-event reconstruction efficiencies are largely independent of the specific tag channel chosen, we will, in the numerical computation of the signal-event yields, include all tag channels for each of which we use the reconstruction efficiencies of the $D^- \to K^+ \pi^- \pi^-$ channel.

Finally we select candidate particles, extract relevant observables from kinematic properties, and apply selection criteria to identify signal events and compute the efficiencies.

\begin{table}[t]
\centering
\begin{tabular}{|c|c|c|}
\hline
selection criteria&cut&efficiency  \\ \hline
$\Delta E_\text{tag}$ & $-0.014\text{~GeV}<\Delta E_\text{tag}<0.013\text{~GeV}$ & $46.10\%$  \\ \hline
 $M_{\text{BC}}$& $1.867\text{~GeV}<M_{\text{BC}}<1.873\text{~GeV}$ & $43.11\%$   \\ \hline
 $\text{PID}_\text{sig}$&  $N_{\pi}^\text{prompt} \ge 1$, $N_{\pi}^\text{displaced} \ge 1$, $N_{e}^\text{displaced} \ge 1$& $3.66\%$  \\ \hline
 $\Delta E_\text{sig}$&select candidate particle& $3.66\%$  \\ \hline
 $p_T$& $p_T>0.1\text{~GeV}$ & $3.07\%$  \\ \hline
\end{tabular}
\caption{The selection criteria and the corresponding cumulative selection efficiencies, for $m_a=1.4\text{~GeV}$ and $m_N=0.5\text{~GeV}$, combining data samples of $c\tau=1$~m, 5~m, and 10~m.}
\label{tab:selection_criteria}
\end{table}

In Table~\ref{tab:selection_criteria} we list the signal-event selection efficiency cutflows for a benchmark point $m_a=1.4$~GeV and $m_N=0.5$~GeV, combining data samples of $c\tau$ = 1~m, 5~m, and 10~m.
On the tag side, we select, in each event, a set of candidate charged particles by comparing $\Delta E_{\text{tag}}= E_{K\pi\pi}-E_{\text{beam}}$, the difference between the energy of the system of the promptly produced charged particles and the beam energy $E_{\text{beam}}=1.8865~\text{GeV}$.
Since the $\Delta E_{\text{tag}}$ distribution exhibits a sufficiently narrow peak, we impose a $\pm 3\sigma$ requirement around its central value, where $1\sigma = 0.0045$~GeV is obtained by fitting the $\Delta E_{\text{tag}}$ distribution with a crystal ball function.
Subsequently, we compute the beam-constrained mass $M_{\text{BC}}=\sqrt{E^2_{\text{beam}}-|\boldsymbol{p}_{D^-}|^2}$ for each candidate combination, and apply a $\pm 3\sigma$ window around the central value $M_{\text{BC}}=1.8697~\text{GeV}$ to the event yield, where $1\sigma = 0.0011$~GeV is attained similarly as for $\Delta E_{\text{tag}}$.
Here, $\boldsymbol{p}_{D^-}$ is the three momentum of all candidate particles for reconstructing the $D^-$-meson in the COM system.

\begin{figure}[t]
    \centering
    \includegraphics[width=0.5\textwidth]{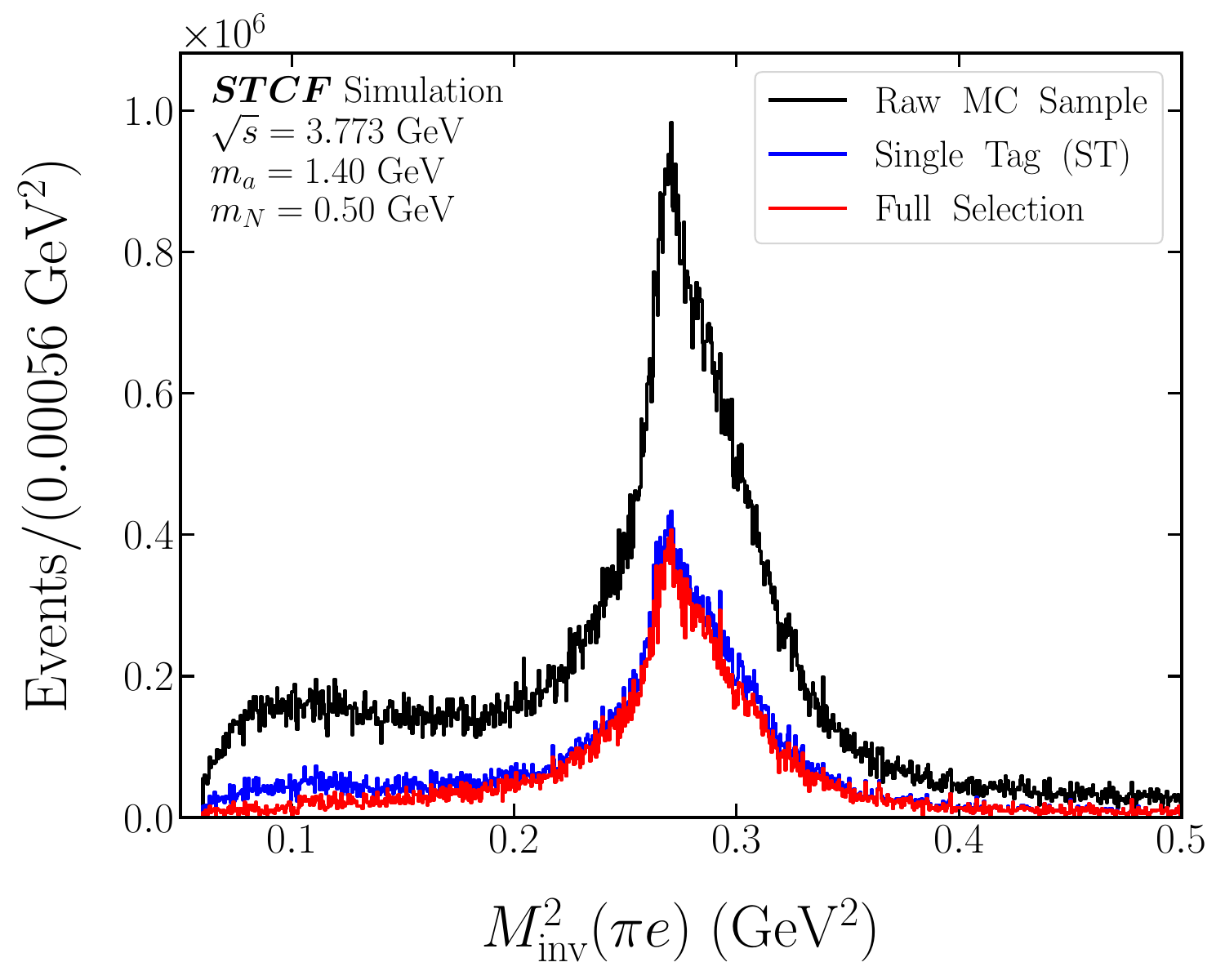} 
    \caption{Distributions of the $\pi e$ invariant mass squared for $m_a=1.4$~GeV and $m_N=0.5$~GeV, at $\sqrt{s}=3.773$~GeV, obtained with OSCAR simulations. The black, blue, and red curves correspond to the raw MC samples, single-tagged samples, and full-selection samples, respectively. Samples of $c\tau=1$~m, 5~m, and 10~m are combined.}
    \label{fig:pie_inv}
\end{figure}

On the signal side, we first select charged particles stemming from DVs by requiring $d_0 > 0.4~\text{cm}$ and $z_0 > 0.6~\text{cm}$, which are the transverse and longitudinal distances of the point of closest approach to the IP, respectively.
This allows to distinguish displaced tracks from candidate particles of the tagging side as well as other prompt particles.
We then require there should be at least one displaced $\pi$ candidate and at least one displaced $e$ candidate, with the $\pi$ and $e$ having opposite electric charges.
We also demand at least one promptly produced $\pi$ with a charge opposite to that of the tag-side $D^-$-meson.
Subsequently, a $\Delta E_{\text{sig}}$ variable, defined as the difference between the energy of the displaced charged particles and the beam energy, is utilized to further select these charged particles and obtain a set of signal-side candidates, but no window requirement around the central value of $\Delta E_{\text{sig}}$ is imposed.
Finally, we require the transverse momentum of both the displaced $\pi$ and displaced $e$ candidates to satisfy $p_T > 0.1~\text{GeV}$, to improve the $\pi/e$-separation capability and the track-reconstruction precision.
The final signal is observed through the distribution of the invariant mass $M^2_{\text{inv}}(\pi e)$ reconstructed from the $\pi e$ pair, as illustrated in Fig.~\ref{fig:pie_inv} for $m_a=1.4$~GeV and $m_N=0.5$~GeV.
We choose not to apply additional selections on $M^2_{\text{inv}}(\pi e)$, since doing so would potentially lead to loss of signal peaks in a real experimental analysis.

As can be observed from Table~\ref{tab:selection_criteria}, after imposing the PID selection criteria on the signal side, the corresponding selection efficiency decreases significantly.
This reduction is mainly driven by the requirement of the displaced pions and electrons identified in front of the MDC.
Specifically, this effect stems from both the decrease in the detector reconstruction efficiency with increasing HNL radial decay position $r$, and the fact that some long-lived HNLs in our event samples do not decay inside the fiducial volume.
Meanwhile, in Fig.~\ref{fig:pie_inv}, the full-selection sample does not exhibit a variation as significant as that shown in Table~\ref{tab:selection_criteria} relative to the single-tag selected sample.
The reason is that, for events passing the single-tag selection but failing the signal-side PID selection, the kinematic variable $M^2_{\text{inv}}(\pi e)$ is often not properly assigned in the analysis program and therefore falls outside the displayed range of the horizontal axis.

In Fig.~\ref{fig:pie_inv} we have combined all event samples of $c\tau=1$~m, 5~m, and 10~m.
The black, blue, and red curves correspond to the raw MC sample, single-tag selected sample (including the $\Delta E_\text{tag}$ and $M_{\text{BC}}$ selections in Table~\ref{tab:selection_criteria}), and the final full-selection sample, respectively.
A peak is observed at $M^2_{\text{inv}}(\pi e)=0.25$~GeV$^2$ corresponding correctly to the HNL mass of 0.5~GeV.
In the low invariant-mass region of the raw MC sample, around $0.1~\text{GeV}^{2}$, the invariant-mass distribution exhibits an anomalous enhancement.
This is mainly because some low-energy hits are reconstructed as fake tracks, which are then incorrectly identified as $e$ or $\pi$ particles.
Importantly, the distributions show that with the cut selections we impose on the event samples, the signal peak around $0.25$~GeV$^{2}$ becomes more pronounced and exhibits a more Gaussian-like shape, with the tails on both sides winding down.
Therefore, with the application of our selection criteria, such misreconstruction is effectively suppressed.

Similar to the approach employed in our phenomenological analysis, we perform a scan over the HNL mass in the range $0.2$--$0.8\text{~GeV}$ with a step size of $0.1~\text{GeV}$.
However, we have considered only three representative values of $c\tau$ approximately matched to the detector scale, i.e.~$1~\text{m}$, $5~\text{m}$, and $10~\text{m}$, in order to ensure sufficient statistics of HNL decays occurring within the fiducial volume.
This choice is motivated by the fact that the reconstruction efficiency $\epsilon_{\text{rec}}(r,m_a,m_N)$ of the signal channel is intrinsically independent of $c\tau$, and depends rather on the ALP mass, the HNL mass, and its radial decay position $r$ within the tracker.
Thus, instead of parameterizing the reconstruction efficiencies in terms of $|V_{eN}|^2$ or $c\tau$, we work with the efficiencies as functions of $r$ which provide a more fundamental characterization and allow to obtain a reliable estimate without simulating prohibitively large event samples.

\begin{figure}[t]
     \centering
     \begin{subfigure}[b]{0.495\textwidth}
         \centering
         \includegraphics[width=\textwidth]{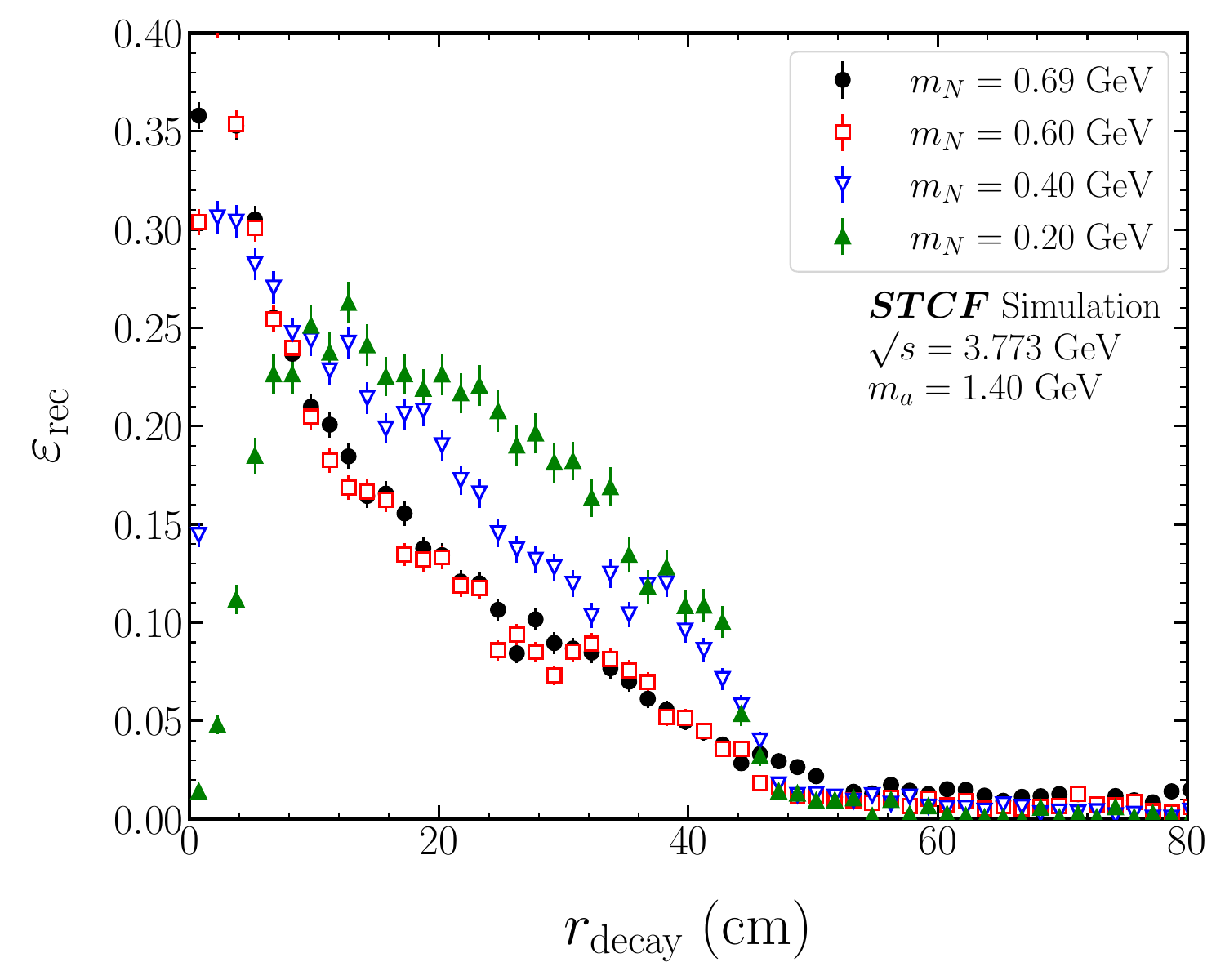}
         \caption{$m_a=1.40~\text{GeV}$}
         \label{fig:recon_effi_1p4}
     \end{subfigure}
     \hfill
     \begin{subfigure}[b]{0.495\textwidth}
         \centering
         \includegraphics[width=\textwidth]{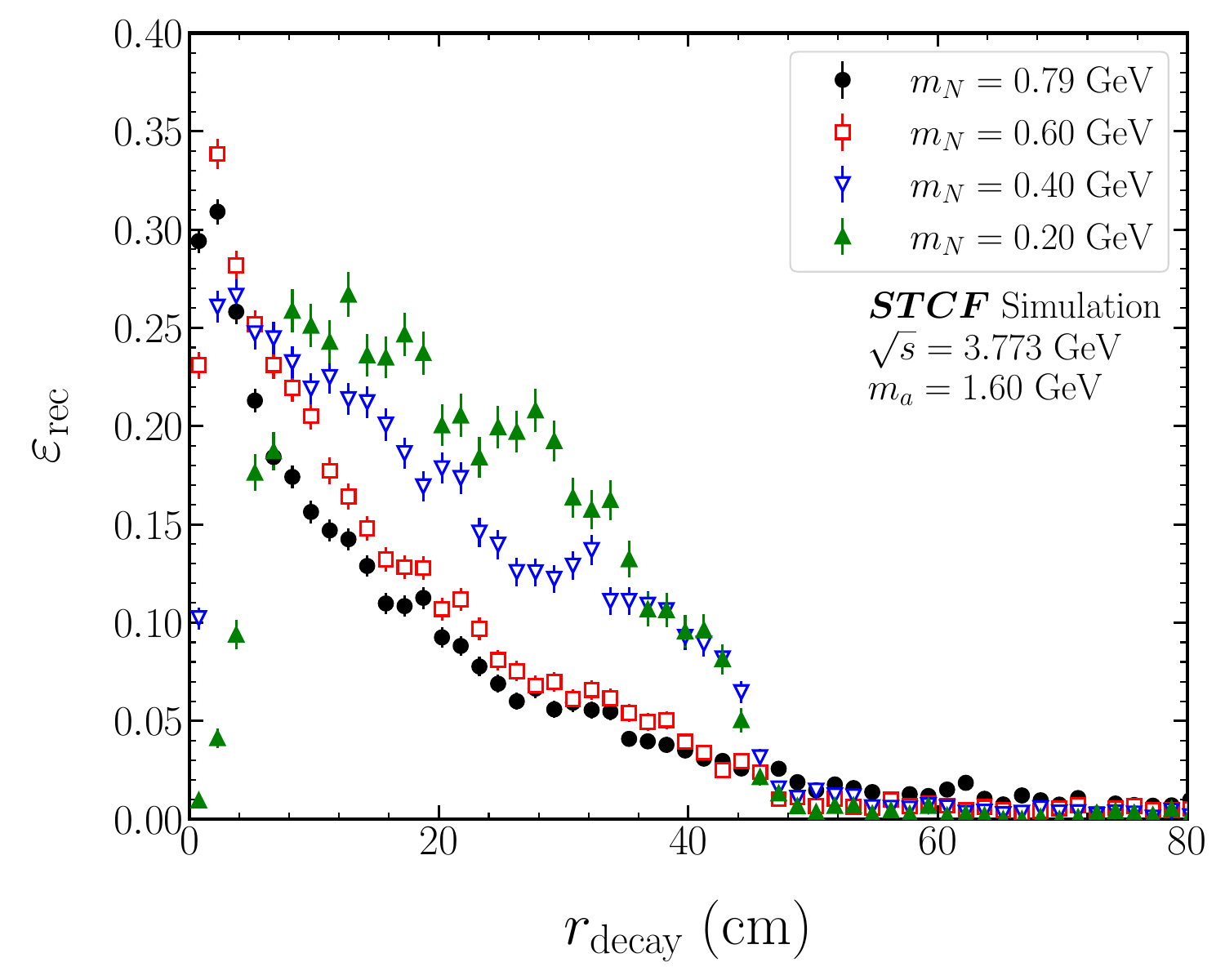}
         \caption{$m_a=1.60~\text{GeV}$}
         \label{fig:recon_effi_1p6}
     \end{subfigure}
     \caption{Reconstruction efficiencies $\epsilon_{\text{rec}}$ as functions of the transverse decay position from the IP $r_{\text{decay}}$, for $m_a=1.4$~GeV (left) and $m_a=1.6$~GeV (right). For each benchmark ALP mass, the efficiencies are shown for representative HNL masses.} 
     \label{fig:recon_effi}
\end{figure}

The reconstruction efficiencies $\epsilon_{\text{rec}}(r,m_a,m_N)$ with respect to the transverse decay position $r_{\text{decay}}$ relative to the IP are plotted in Fig.~\ref{fig:recon_effi}, for representative values of $m_a$ and $m_N$.
To obtain Fig.~\ref{fig:recon_effi}, we divide the fiducial volume into bins along the radial direction $r$ with a width of $1.5$~cm, covering the range $0\text{~cm} < r < 85$~cm, i.e.~the region between the IP and the edge of MDC.
For each bin, we compute the ratio of the number of reconstructed signal events to the MC truth-level event number of HNL decays within this unit volume, defined as the signal-event reconstruction efficiency at the bin center and shown as a point in Fig.~\ref{fig:recon_effi}.

In this numerical simulation we use the benchmark $c\tau$ values of $\mathcal{O}(1)$~m which are in the detector length scale, to obtain large statistics of HNL decays within the fiducial volume.
However, this way the probability of \textit{both} HNLs decaying inside the fiducial volume is non-negligible.
Thus, when evaluating the reconstruction efficiency for one of the two HNLs produced in ALP decays, we require the other HNL to decay outside the whole fiducial volume, to avoid contamination from decays occurring outside the tested bin region.
This requirement is consistent with Eq.~\eqref{eqn:epsilon} used in our phenomenological analysis.
Note that for computing this efficiency we have selected all the events of HNL decays within the unit volumes that have passed the tag-side selection criteria including the $\Delta E_\text{tag}$ and $M_{\text{BC}}$ ones.

We observe that the reconstruction efficiencies start with relatively low values at small $r_{\text{decay}}$, reach a maximum at intermediate radii, and then gradually degrade for larger $r_{\text{decay}}$.
The suppression at small $r_{\rm decay}$ originates from the requirement of identifying both the electron and the pion as displaced tracks.
For decays occurring close to the IP, the electron and pion can easily fail the requirements of $d_0>0.4~\text{cm}$ and $z_0>0.6~\text{cm}$ on both of them.
Moreover, lighter HNLs are typically more boosted and their decay products tend to be more collimated along the HNL flight direction, thus further decreasing the probability of passing these requirements.
On the other hand, the reconstruction efficiencies weaken at large distances from the IP because of the degrading tracking capability of the tracker.

The detector-level reconstruction efficiency $\epsilon_{\text{rec}}(r,m_a,m_N)$ is more realistic and reliable than the linear assumption $L(r)$ in Eq.~\eqref{eqn:linear_efficiency}.
Replacing $L(r)$ in Eq.~\eqref{eqn:prob} with $\epsilon_{\text{rec}}(r,m_a,m_N)$ allows for extrapolating the detector-level signal acceptances to any $c\tau$ values.
This approach has been previously adopted in, for example, a Belle search~\cite{Belle:2022tfo}.

The results presented above are obtained from the full detector OSCAR simulation of only the signal MC samples.
We have also analyzed inclusive MC samples of $D^+D^-$ pairs, including all SM decay channels, corresponding to approximately $\mathcal{O}(10^7)$ events to obtain a qualitative understanding of the background.
The results are consistent with the expectations in Sec.~\ref{sec:truthlevel}: by applying the same selection criteria as those used in the signal MC simulation, the physical backgrounds can be effectively suppressed, and the remaining background is dominated by fake tracks arising from misreconstruction of detector responses, as well as combinatorial backgrounds.
These backgrounds can be further reduced by performing vertex fitting on the final-state tracks and by employing machine-learning techniques \cite{Jia:2025ufk, Zhai:2025qke}, while maintaining the original reconstruction efficiency.
It should be noted that, owing to computational resource limitations, only a limited inclusive MC sample is analyzed, which is insufficient to draw final conclusions about the background level corresponding to an integrated luminosity of $1~\text{ab}^{-1}$.
A more comprehensive background study and further optimization of the selection criteria are beyond the scope of the present work and are left for future experimental studies.

Based on the discussions in the previous and present sections, we will assume vanishing background levels and derive the contours corresponding to 3 signal events, representing the sensitivity reach at the $95\%$ confidence level (CL).

\section{Numerical results}\label{sec:results}

\begin{figure}[t]
    \centering
    \includegraphics[width=0.6\textwidth]{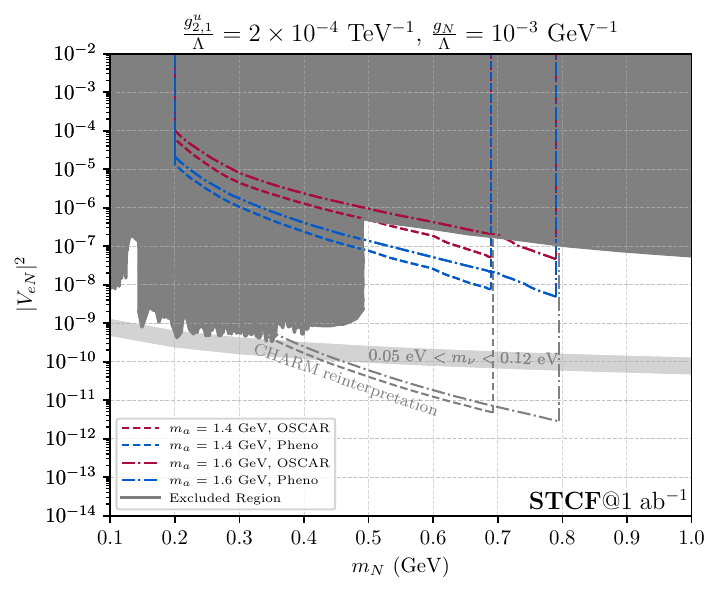} 
    \caption{Sensitivity reach of STCF with an integrated luminosity of 1~ab$^{-1}$ to the long-lived HNLs from the prompt decays of the ALP stemming from $D^+$-meson decays. We fix $g^u_{2,1}/\Lambda$ at $2\times 10^{-4}$~TeV$^{-1}$ and $g_N/\Lambda$ at $10^{-3}$~GeV$^{-1}$. The shown results are for benchmark values of $m_a=1.4$~GeV (dashed) and 1.6~GeV (dot-dashed). The red and blue curves are obtained with the full OSCAR simulation and a quick truth-level simulation, respectively. The existing bounds, obtained at PIENU~\cite{PIENU:2017wbj}, KENU~\cite{Bryman:2019bjg}, CHARM~\cite{CHARM:1985nku}, NA62~\cite{NA62:2020mcv,NA62:2025csa}, T2K~\cite{T2K:2019jwa}, and BEBC~\cite{Barouki:2022bkt}, are shown in dark gray, and the approximate CHARM-search reinterpretation results, obtained under simplified assumptions on the production and kinematic distributions, are displayed in dot-dashed gray lines.}
    \label{fig:sensitivity}
\end{figure}

We present the sensitivity reach of STCF with 1~ab$^{-1}$ to the long-lived HNLs as a function of $m_N$ in Fig.~\ref{fig:sensitivity}.
We have fixed the couplings $g^u_{2,1}/\Lambda$ at $2\times 10^{-4}$~TeV$^{-1}$ and $g_N/\Lambda$ at $10^{-3}$~GeV$^{-1}$ ensuring the promptness of the ALP.
For the two benchmark values of $m_a$ we consider, 1.4~GeV (dashed curves) and 1.6~GeV (dot-dashed curves), we show the numerical results obtained with the quick phenomenological analysis (blue) and the realistic OSCAR-simulation analysis (red).
The realistic sensitivities are weaker than those of the simple phenomenological analysis.
This is mainly because the position-dependent efficiencies estimated in the OSCAR-based analysis are somewhat lower than those associated with the linear function, Eq.~\eqref{eqn:linear_efficiency}, employed in the simple analysis.

The existing bounds obtained at PIENU~\cite{PIENU:2017wbj}, KENU~\cite{Bryman:2019bjg}, CHARM~\cite{CHARM:1985nku}, NA62~\cite{NA62:2020mcv,NA62:2025csa}, T2K~\cite{T2K:2019jwa}, and BEBC~\cite{Barouki:2022bkt} are shown in the dark-gray region, while the approximate reinterpretation results of the CHARM search~\cite{CHARM:1985nku}, obtained under simplified assumptions on the production mechanism and kinematic distributions of the HNLs, are displayed in dot-dashed gray lines.
Furthermore, we plot a band in light gray corresponding to the parameter region targeted by the type-I seesaw mechanism, for active-neutrino masses between 0.05~eV and 0.12~eV.
The lower and upper bounds on the active-neutrino masses originate from experiments searching for neutrino oscillations~\cite{Canetti:2010aw} and from cosmological observations~\cite{Planck:2018vyg}, respectively.

These results show that STCF is expected to probe the mixing parameter $|V_{eN}|^2$ about one-to-two orders of magnitude beyond the existing bounds for $m_N$ between 0.5~GeV and 0.8~GeV,
although they do not yet reach the theoretical parameter space predicted by the type-I seesaw mechanism (the light-gray band in Fig.~\ref{fig:sensitivity}).
We note, however, that the approximate reinterpretation of the CHARM search~\cite{CHARM:1985nku} in the present ALP--HNL framework is subject to model dependence, as it relies on assumptions about the production mechanism and kinematic distributions of the HNLs. 
In particular, the presence of an intermediate ALP modifies both the production and decay kinematics of the HNLs, which can affect the acceptance of beam-dump experiments. 
Therefore, this reinterpretation shown here should be regarded as indicative rather than a definitive exclusion. A dedicated recast including full detector simulation would be required for a robust comparison, which is beyond the scope of the present work. 
Refs.~\cite{Wang:2024prt,Wang:2024mrc} investigated the expected signals of the ALP-HNL model at other future experiments such as LHC far detectors and SHiP, as well as the ongoing experiment Belle~II, providing exclusion limits on $|V_{eN}|^2$.
Ref.~\cite{Wang:2024prt} studied a benchmark scenario with an ALP produced in $B$-meson decays (instead of $D$-meson decays) and Ref.~\cite{Wang:2024mrc} investigated benchmark scenarios including ours.
However, even in the latter case, the considered ALP mass values are different from ours.
Therefore, our study should be regarded as being highly complementary in terms of the parameter space coverage.
For these reasons, we do not overlap such existing results explicitly in our sensitivity plot; nevertheless, it can be inferred from the results given in Ref.~\cite{Wang:2024mrc} that for our benchmark scenario the future LHC far-detector and the SHiP experiments are expected to probe values of $|V_{eN}|^2$ orders of magnitude below those to which STCF is sensitive.
In this context, STCF provides a complementary probe of this scenario, with a clean experimental environment and well-controlled kinematics at threshold.

For smaller values of $g^u_{2,1}/\Lambda$ the sensitivities are expected to weaken.
Quantitatively, if $g^u_{2,1}/\Lambda$ is reduced by a factor of, say, 10, the sensitivity reach to $|V_{eN}|^2$ is expected to become weaker by 100.

\section{Conclusions}\label{sec:conclu}

In this work, we have considered an ALP coupled to the charm and up quarks off-diagonally, and to a pair of long-lived Majorana HNLs.
We have performed both fast and dedicated MC simulations in order to assess the search prospect of the long-lived HNLs at the proposed Super Tau--Charm Facility.

We have focused on the COM energy mode of $\sqrt{s}=3.773$~GeV at STCF and the $D^+\to \pi^+ a$ and $a\to NN$ decays.
The ALP $a$ is on-shell and decays promptly and exclusively into a pair of $N$'s.
With both a simple phenomenological analysis and a dedicated reconstructed-level analysis, we determine the sensitivity reach to the active-neutrino-HNL mixing parameter $|V_{eN}|^2$ as a function of the HNL mass, where we have fixed the decay branching ratio of $D^+\to \pi^+ a$ at the experimentally allowed upper bound by assuming a sufficiently small value of $g^u_{2,1}/\Lambda$.

Our numerical results show that STCF can probe values of $|V_{eN}|^2$ up to about one-to-two orders of magnitude below the existing bounds, while an approximate reinterpretation of a CHARM search is found to surpass the STCF sensitivities.
A full and dedicated procedure of recast and reinterpretation would be required to establish such exclusion bounds, which is, however, out of the scope of the present work.
Moreover, for benchmark scenarios with the HNL mixed purely with $\nu_\mu$ ($\nu_\tau$) instead of $\nu_e$, the sensitivities are expected to be comparable (weaker), considering the similar (distinct) HNL decay products.

Finally, we comment that our analysis is readily generalizable to additional theoretical scenarios such as a scenario with both a new scalar and a dark photon, and the dark axion portal scenario.

\acknowledgments
We would like to thank, in alphabetical order, Xu Han, Qingyuan Huang, Minghui Liu, and Mingyi Liu for useful discussions.
We acknowledge support by the National Natural Science Foundation of China under Grants No.~12475106, No.~12505120, and No.~12341504, the Fundamental Research Funds for the Central Universities under Grant No.~JZ2025HGTG0252, and the CAS Youth Team Program under Contract No.~YSBR-101.

\bibliography{refs}

\end{document}